\documentstyle[aps]{revtex}
\begin{document}
\draft

\title{Composition of the nuclear periphery from antiproton absorption}

\author{P. Lubi\'nski, J. Jastrz\c{e}bski, and A. Trzci\'nska}
\address{Heavy Ion Laboratory, Warsaw University, PL-02-093 Warsaw, Poland}
\author{W. Kurcewicz}
\address{Institute of Experimental Physics, Warsaw University, PL-00-681 
Warsaw, Poland} 
\author{F. J. Hartmann, W. Schmid, and T. von Egidy}
\address{Physik-Department, Technische Universit\"at M\"unchen, 
D-85747 Garching, Germany}
\author{R. Smola\'nczuk and S. Wycech}
\address{So{\l}tan Institute for Nuclear Studies, PL-00-681 Warsaw, Poland}

\date{\today}
\maketitle

\begin{abstract}

Thirteen targets with mass numbers from 58 to 238 were irradiated with the 
antiproton beam from the Low Energy Antiproton Ring facility at CERN leading 
to the formation of antiprotonic atoms of these heavy elements. The antiproton 
capture at the end of atomic cascade results in the production of more or less 
excited residual nuclei. The targets were selected with the criterion that 
both reaction products with one nucleon less than the proton and neutron 
number of the target are radioactive. The yield of these radioactive products 
after stopped-antiproton annihilation was determined using gamma-ray 
spectroscopy techniques. This yield is related to the proton and neutron 
density in the target nucleus at radial distance corresponding to the 
antiproton annihilation site. The experimental data clearly indicate the 
existence of a neutron rich nuclear periphery, a ``neutron halo'', strongly 
correlated with the target neutron separation energy B$_{n}$, and observed for 
targets with B$_{n}$ $<$ 10 MeV. For two target nuclei, $^{106}$Cd and 
$^{144}$Sm, with larger neutron binding energies a proton rich nuclear 
periphery was observed. Most of the experimental data are in reasonable 
agreement with calculations based on current antiproton-nucleus and 
pion-nucleus interaction potentials and on nuclear densities deduced with the 
help of the Hartree-Fock-Bogoliubov approach. This approach was, however, 
unable to account for the $^{106}$Cd and $^{144}$Sm results.

\end{abstract}



\section{Introduction}

There is a growing experimental evidence \cite{wilkinson59,davis67,burhop67,%
burhop69,nolen69,korner71,burhop72,bugg73a,bugg73b,leon74,bugg75,%
jastrzebski93a,lubinski94,adams96} that the outer periphery of many stable 
isotopes of heavy elements is composed predominantly of neutrons. Although 
this is in agreement with simple \cite{bethe70} as well as sophisticated 
\cite{negele70} nuclear models, its recognition could only rarely be found 
in the literature \cite{bohr69,jackson75}. Indeed, the discussions of 
differences between matter and charge distributions were generally limited 
to the comparison of the corresponding mean square radii \cite{angeli86,%
gambhir90,baran95}, a quantity much easier accessible to the experiment 
\cite{batty89,krasznahorkay91} than the composition of the nuclear 
stratosphere (with the nucleon densities two or three orders of magnitude 
smaller than the central density). Recently, however, the interest in the 
composition of the outermost nuclear periphery was largely increased when 
it was realised \cite{dobaczewski96a,dobaczewski96b,hamamoto96} that the 
asymptotic behaviour of the nuclear wave function may govern a number of 
phenomena which are expected in experiments with radioactive beams. 
In particular, a non-homogeneous distribution of the extra-neutrons 
(or extra-protons) in nuclei far from stability may lead to the existence 
of a marked neutron (proton) halo. Such halos would certainly manifest 
themselves e.g. in low energy transfer reactions induced by radioactive 
beams or, as recently demonstrated \cite{takahashi97}, will strongly 
influence the fusion probability. Before detailed studies of reactions 
induced by neutron rich or neutron poor projectiles are undertaken it is 
of considerable interest to understand how the nuclear periphery of stable 
nuclei changes its properties as a function of isospin. If this is well 
described by existing theories, one can at least hope that an extrapolation 
to more exotic nuclei/projectiles (with much larger or smaller isospin 
values than those for stable ones) will be not too far from the reality.

This example notwithstanding, the composition and extent of the nuclear 
periphery are evidently of a considerable interest by themselves. Governed 
by the asymptotic behaviour of the nuclear wave function they can be 
a sensitive testing ground for nuclear models going beyond harmonic oscillator 
boundaries.

Experimentally, the studies of the nuclear periphery are evidently facilitated 
if the probes used interact very strongly with nucleons leading to clear 
signals even from the diluted nuclear stratosphere. Indeed, a substantial part 
of our present information on the extent and composition of the outer nuclear 
periphery was obtained via the formation of hadronic (pionic, kaonic and 
antiprotonic) atoms. Two types of experiments can be performed. In the first 
one the X-ray spectra of these ``exotic`` atoms are investigated 
\cite{backenstoss74,kohler86,garciarecio92,batty95} and in particular 
the characteristics of the last X-ray transition which can be observed before 
hadron - nucleus interaction prevents further X-ray emission are studied. 
These characteristics are governed by the hadron - nucleus interaction 
potential depending in turn on the matter density where the interaction 
occurred. In the second type of experiments also involving the formation of 
hadronic atoms, the characteristics of the hadron annihilation products - 
mesonic \cite{davis67,bugg73a,bugg73b} or nuclear \cite{jastrzebski93a} - 
are studied. The formation and nature of these products depend not only on 
the nuclear density distribution but are also related to the neutron to proton 
density ratio.

We have recently reported \cite{jastrzebski93a,lubinski94} a new experimental 
study of the nuclear periphery using the formation of antiprotonic atoms. 
In this study the nuclear products of antiproton annihilation on a peripheral 
proton or neutron were identified using simple radiochemical methods. It was 
shown that the outer periphery of the heaviest isotopes of all elements 
studied is composed to a large extent of neutrons. In addition, a strong 
correlation between the neutron to proton density ratio and the neutron 
binding energy was found.

In the present paper we give a more detailed description of this experiment 
together with some new data.

\section{THE METHOD}

After the pure and intense antiproton beam from the LEAR 
facility at CERN \cite{gastaldi79,lefevre87,jones87} became available the 
yield of the radioactive products formed after stopped antiproton - target 
interaction could be investigated. Such studies were undertaken more than 
ten years ago \cite{moser86,moser89,egidy90} and were continued thereafter 
\cite{jastrzebski93b,jastrzebski95,lubinski98}. Their main objective, similar 
to that for radiochemical work conducted for decades with protons, heavy ions 
or pions (see e.g. Ref. \cite{cumming74,pienkowski91,haustein78}) is the study 
of the energy transfer to the target nucleus after the interaction. The 
unusual character of the antimatter projectile, its decay characteristics, 
the hope for some exotics add to the interest of such investigations.

Radiochemical experiments with stopped antiprotons were very simple when using 
the LEAR facility. The extracted extremely pure and monoenergetic antiproton 
beam was traversing a thin scintillator counter (determining the number of 
antiprotons) and was impinging on the target thick enough to stop antiprotons. 
If the antiproton energy was too high a beam energy degrader was used in front 
of the target. Preferably the target thickness should be larger than the 
antiproton pathlength straggling but this was not always possible with 
expensive, isotopically separated targets. 

An antiproton entering the target material looses energy by interaction with 
atomic and conduction electrons. When its kinetic energy is comparable to the 
ionisation energy of the target atoms the antiproton replaces one of the 
atomic electrons and occupies a high-{\it n} orbit of this atom. 
After the capture an electromagnetic cascade develops through 
lower, empty ({\it n},{\it l}) states. Spatially, due to the large 
antiprotonic mass, a good fraction of this cascade occurs inside the 
electronic cloud. The energy released by the cascading antiproton is in 
the beginning of the cascade (when the antiproton is in high {\it n} orbits, 
close to the electron orbits) taken away by Auger electrons. For lower 
antiprotonic orbits the emission of electromagnetic radiation becomes 
important and this X-ray emission dominates for the lowest antiprotonic 
transitions. The statistical population of the {\it l}-substates at the moment 
of capture \cite{eisenberg61,leon77a,leon77b} and the selection rules for 
electromagnetic E1 transitions favour 
the population of highest-{\it l} levels (={\it n}-1) at the end of the 
cascade. Cascading between these circular orbits the antiproton eventually 
enters the nuclear periphery where, after an encounter with a proton or with 
a neutron, it annihilates.

As a result, in more than 95\% of the annihilations charged and neutral pions 
are formed with a multiplicity ranging from two to eight and with an average 
value equal to five. Some of those pions may enter the nuclear volume 
initiating an intra-nuclear cascade \cite{clover82,iljinov82,cugnon85} and 
heating the nucleus which subsequently emits neutrons, light charged 
particles, intermediate mass fragments or undergoes fission. The development 
of these processes evidently critically depends on the amount of transferred 
energy which in turn is related to the number, n{\normalsize $_{int}$}, of 
pions directed towards the inner nucleus and interacting inelastically with 
it. (Due to the strong pion - nucleus potential this interaction is generally 
followed by pion absorption).

The earliest experimental determination of $<$n$_{int}>$ was 
presented by Bugg et al. \cite{bugg73b}. Subsequent experiments are discussed 
and analysed by Cugnon et al. \cite{cugnon89} and the most recent data are due 
to Polster et al. \cite{polster95}. As it could be expected the average value 
of n$_{int}$ depends on the target mass number 
A$_{t}$ and $<$n$_{int}>$ is about one for 
A$_{t} \approx$ 200  and less for lighter target nuclei.

This relatively small value of the average number of pions interacting with 
the target nucleus immediately indicates that the annihilation events with 
n{\normalsize $_{int}$} = 0 should occur with substantial probability. 
This is, indeed, confirmed by intranuclear-cascade calculations 
\cite{cugnon87} which predict that these ``void cascade events'' appear 
in 10-20\% of the annihilations. After such events cold residual nuclei are 
formed with a mass equal to the target mass decreased by the mass of one 
nucleon which participated in the annihilation process. Experimentally, such 
products with mass (A{\normalsize $_{t}$} - 1) were clearly observed with 
a large yield in the radiochemical studies of the stopped antiproton 
interaction with nuclei \cite{moser89,jastrzebski93b}.

In the Nuclear Chart one can find a number of target nuclides for which both 
neighbouring products with one nucleon less than those in the target with mass 
number A{\normalsize $_{t}$}=N{\normalsize $_{t}$}+Z{\normalsize $_{t}$} are 
radioactive. For such target nuclei the yield ratio of the 
(N{\normalsize $_{t}$} - 1) to the (Z{\normalsize $_{t}$} - 1) products after 
the annihilation, easily determined using classical nuclear spectroscopy 
methods, will give information on the neutron to proton ``concentration 
ratio'' in the region where annihilation occurred.

The information on the annihilation site can also be obtained from such 
a simple radiochemical experiment providing the fraction, 
Y(A{\normalsize $_{t}$} - 1), of annihilation events leading to the 
(A{\normalsize $_{t}$} - 1) products is determined. This fraction depends on 
the product of the antiproton annihilation probability, W(r), where r is the 
radial distance from the center of the nucleus, with P{\normalsize $_{miss}$}
(r). The last quantity expresses the ``missing probability'' 
\cite{jastrzebski93a} that all pions created during the antiproton 
annihilation miss the inner nucleus, leading to n{\normalsize $_{int}$} = 0. 
P{\normalsize $_{miss}$}(r) evidently increases with the radial distance, r. 
Therefore qualitatively, the larger the Y (A{\normalsize $_{t}$} - 1) value is 
the farther from the nuclear center the neutron to proton ratio is tested.

The experiment determines the yield ratio of (N{\normalsize $_{t}$} - 1) 
to (Z{\normalsize $_{t}$} - 1) products and the fraction of events leading to 
(A{\normalsize $_{t}$} - 1) nuclei. To interpret the data and to know at which 
radial distance the method tests the composition of the nuclear periphery - 
the recourse to the theory is necessary. Recent calculations indicate 
\cite{wycech96} that almost independently of the target mass the antiproton 
annihilation probability W(r) has its maximum value at distances about 2 fm 
larger than the half-density radius. Its convolution with 
P{\normalsize $_{miss}$}(r) locates the maximum of n{\normalsize $_{int}$} = 0 
events at a distance even about 1 fm larger, making the method sensitive to 
the composition of the outermost nuclear periphery with nuclear densities of 
about 10{\normalsize $^{-2}$} - 10{\normalsize $^{-3}$} of the central 
density. We shall come back to these questions at the end of this paper.

In order to be able to compare the experimental data for any measured target, 
independently of its neutron to proton ratio, we introduce the neutron halo 
factor defined as 
\[f^{periph}_{halo} = \frac{N(\bar{p},n)}{N(\bar{p},p)}\frac{Im(a_{p})}%
{Im(a_{n})}\frac{Z_{t}}{N_{t}}\]
where $N(\bar{p},n)/N(\bar{p},p)$ is the ratio of produced A$_{t}$ - 1 nuclei, 
and a$_{n}$ and a$_{p}$ are the $\bar{p}$-n and $\bar{p}$-p scattering 
amplitudes, respectively \cite{leon74}. The ratio 
Im(a$_{p}$)/Im(a$_{n}$) = 1/R$_{np}$ 
accounts for the ratio of annihilation probabilities. A similar halo factor 
was previously introduced by Bugg et al. \cite{bugg73b} who investigated the 
composition of the nuclear periphery determining the characteristics of 
mesonic products of antiproton annihilation. In his method all annihilation 
events contributed to the halo factor, whereas only the most peripheral events 
are of importance in the present method, so the superscript ``periph'' is used 
here.

Neglecting the corrections mentioned below, the halo factor accounts 
essentially for the enhancement of the neutron or proton concentration over 
the normal one (i.e. that, which reflects the N/Z target ratio) in the tested 
nuclear region. A halo factor higher than one would indicate an increased 
neutron concentration (``neutron halo'') whereas f$^{periph}_{halo} <$ 1 would 
correspond to an enhanced proton density at the nuclear periphery.

There are, however, corrections which have to be discussed. The above 
considerations assume that all events in which n{\normalsize $_{int}$} = 0 
lead to cold nuclei with mass number (A{\normalsize $_{t}$} - 1). This is not 
necessarily the case. Microscopically, the nuclear periphery may contain some 
amplitudes of states bound more strongly than the nucleon separation energies 
or fission barrier of these (A{\normalsize $_{t}$} - 1) nuclei. In such a case 
the antiproton annihilation on these deeply bound states would lead to high 
rearrangement energies \cite{hodgson81} of the (A{\normalsize $_{t}$} - 1) 
products, which would subsequently emit particles or fission. A simple 
Thomas-Fermi model estimate \cite{blocki93} of this effect or more detailed 
calculations within the Hartree-Fock-Bogoliubov model reported in Ref. 
\cite{wycech96} show that this can affect the halo factor by no more than 30\%.

Another effect which merits consideration is the possibility of the pion 
charge exchange process which could change the final population of 
(Z{\normalsize $_{t}$} - 1, N{\normalsize $_{t}$}) and (Z{\normalsize $_{t}$},
N{\normalsize $_{t}$} - 1) nuclei. The yield of this process was determined 
experimentally and is discussed in Sec. IVc.

Ending this section about the experimental method one should mention the 
article by Bloom et al. \cite{bloom72}, known to us well after the first 
presentation \cite{jastrzebski93a,lubinski94} of the method described here. 
In this article the idea of looking on nuclear (rather than mesonic) signals 
when investigating the composition of the nuclear periphery with exotic atoms 
was already suggested. To the best of our knowledge these suggestions were 
never realized in the way proposed there.

\section{EXPERIMENTAL TECHNIQUE}

The experimental results which are presented in this paper were gathered 
during five irradiation series from 1991 till 1996. The antiproton beam 
momenta delivered by the LEAR facility are listed in Table\ \ref{ranges}.

A similar set-up was employed for the experiments with 200 and 310 MeV/c 
antiprotons. After passing through a 18.5 mg/cm{\normalsize $^{2}$} beryllium 
window at the end of the beam tube and after traversing approximately 10 cm of 
air, the antiprotons encountered the first scintillation counter S1 
($\phi$ = 15 mm, thickness 10 mm) with a central hole of 7 mm. This counter, 
used also for the beam focusing, served as an active diaphragm for the 
antiproton counting during the target irradiation.

A variable thickness plastic beam energy degrader was placed immediately after 
the S1 counter. Its selected thickness was calculated to insure that the 
antiproton beam, after traversing the second scintillation counter S2 
($\phi$ - 7 mm, thickness 2.5 mm) and a few Al foils (with a total thickness 
of about 80 mg/cm$^{2}$), stopped approximately in the middle of 
the irradiated target. Similar Al foils were also placed behind the target. 
The $^{24}$Na activity produced by antiprotons stopped in the 
backward and forward Al foils was used to determine the fraction of the 
antiprotons which triggered the S2 counter but were not stopped inside the 
target material. This was due to the fact that the target thickness (indicated 
in Table\ \ref{targets}) was generally smaller than the antiproton pathlength 
straggling (shown in Table\ \ref{ranges} and in Fig.\ \ref{fig1}) induced 
during the energy degradation by the moderator, S2 counter and target 
material. The yield for producing {\normalsize $^{24}$}Na in stopped 
antiproton interaction with Al target was determined by independent 
measurements. The average value found was 19.5$\pm$1.0 {\normalsize $^{24}$}Na 
nuclei produced per 1000 \={p}. The pathlength straggling shown in 
Table \ \ref{ranges} was determined by activation analysis, 
measuring the activity induced by stopped antiprotons in the target composed 
of a stack of thin foils.  

Depending on the half-life of the (A{\normalsize $_{t}$} - 1) reaction 
products, each target stack was bombarded by a short (10-15 min) or long 
(80-90 min) antiproton ``spill'', totalizing to about 5$\times 10^{8}$
 - 10{\normalsize $^{9}$} particles. Generally more than a half of these 
antiprotons were stopped by the target material, whereas the rest reacted with 
Al or other target components (glue, oxygen). In Table\ \ref{targets} the 
targets used in the present work are listed together with the integrated 
intensities of antiprotons stopped in the target nuclei.

A slight modifications of the above described irradiation conditions were 
introduced for the last run, where the antiprotons with the lowest momentum 
were available. During this run the S1 counter with $\phi$ = 8 mm central hole 
and 1 mm thickness preceded the S2 counter with a 100 $\mu$m scintillator. 
Both these counters were placed in a light-tight chamber filled with He gas to 
decrease both slowing down and scattering of the low-energy beam. Al windows 
of 12 $\mu$m thickness were put at entrance and exit of the He chamber. 
Eventually, the 6.0 MeV energy delivered by the LEAR facility was degraded to 
2.8 MeV, with only a rather small energy straggling, as indicated in 
Table\ \ref{ranges}. 
This beam was impinging on the target material, with this time large Al foils 
placed only behind the target to control the beam scattering off the targets 
(diameter 10 mm).

Shortly after the end of the irradiation (about 2 min in the case of short 
lived (A{\normalsize $_{t}$} - 1) products) the gamma-ray counting was started 
using the HPGe coaxial detectors of about 10\ $\div$ 30\ \% relative 
efficiency and with an energy resolution slightly below 2 keV for the 
{\normalsize $^{60}$}Co 1333 keV transition. In some cases, when the low 
energy gamma-rays were of interest, a planar HPGe detector with an energy 
resolution of about 600 eV at 122 keV energy was also employed. The gamma-ray 
counting was generally continued for a few days at CERN and later, for many 
months in Warsaw, where in addition a 60\% HPGe detector was used. The 
production yield of (A{\normalsize $_{t}$} - 1) nuclei was deduced from the 
absolute intensities of their characteristic gamma lines (followed for a few 
half-lives) after corrections for the decay during and after the irradiation. 
To this end the antiproton beam intensity as a function of time was always 
carefully monitored. Generally half-lives, energies and branching ratios were 
taken from the most recent Nuclear Data compilations, available via internet 
connection. In some cases also Ref. \cite{browne86} was used.

For some targets the total number of antiprotons stopped in the target 
material was additionally determined by integrating the mass yield 
distribution of heavy reaction residues. These distributions were gathered 
from the yield of the radioactive reaction products, as previously described 
\cite{jastrzebski93b,jastrzebski95}. It was assumed that for non-fissile 
targets one antiproton stopped in the target material produces one heavy 
reaction residue. The total number of antiprotons determined in this way was 
in general within 10-15\% equal to the number obtained using the indications 
of S2 counter with the corrections described above. Figure\ \ref{fig2} shows 
as an example the mass yield distribution obtained for the $^{130}$Te target. 
More information on experimental details and evaluation procedures 
pertaining to each particular target is presented in Refs. 
\cite{lubinski97a,lubinski97b}.

\section{EXPERIMENTAL RESULTS} 

\subsection{Peripheral neutron to proton density ratio} 

The final results, used to determine the peripheral neutron to proton density 
ratios, are presented in Table\ \ref{results}. These data supersede those 
previously 
published by Lubi\'nski et al. \cite{lubinski94}. The differences are in some 
cases due to more recent information on the decay properties of the 
(A{\normalsize $_{t}$}-1) nuclei and, in other cases, to the inclusion of 
results gathered during the 1995 and 1996 runs. The second, third and fourth 
columns of this Table give the absolute production yield of nuclei one mass 
unit lighter than the target mass. The halo factor, f$^{periph}_{halo}$,
presented in the last column of this Table was obtained from the measured 
ratio of the produced (N{\normalsize $_{t}$}-1) nuclei to the 
(Z{\normalsize $_{t}$}-1) ones (column 5) after correction for 
the target Z{\normalsize $_{t}$}/N{\normalsize $_{t}$} ratio and for the ratio 
of the antiproton annihilation probabilities on a neutron to that on a proton, 
R{\normalsize $_{np}$}. Following Bugg et al. \cite{bugg73b} the value of the 
last ratio was taken to be equal to 0.63, in agreement with Ref. \cite{wade76}.
 Its error is not included in the errors of f$^{periph}_{halo}$ in 
Table\ \ref{results}.

The data from Table\ \ref{results} are also shown in Figs. 3-5. 
Figure\ \ref{fig3} presents the halo 
factor as a function of the target neutron separation energy. The negative 
correlation, previously observed \cite{lubinski94} for a smaller data sample 
is confirmed. The unusually large error for the {\normalsize $^{160}$}Gd 
target is due to the poorly known absolute transition intensities in the decay 
of (A{\normalsize $_{t}$}-1) products \cite{nds94} of this target. The largest 
value of the halo factor is obtained for the {\normalsize $^{176}$}Yb target 
and is discussed below. For two investigated nuclei, {\normalsize $^{106}$}Cd 
and {\normalsize $^{144}$}Sm, the measured halo factor is substantially 
smaller than one. As our assumed value of the ratio 
Im(a{\normalsize $_{n}$})/Im(a{\normalsize $_{p}$}) is probably the lowest 
acceptable (see the discussion in \cite{wycech96}) - these results clearly 
indicate the proton rich atmosphere of these two nuclei. (This would remain 
true for {\normalsize $^{144}$}Sm even if one assumes the value 
R{\normalsize $_{np}$} = 0.48 as obtained in the {\normalsize $^{4}$}He 
experiment \cite{balestra89}). The systematics presented in Fig.\ \ref{fig3} 
shows that 
the nuclei with smaller neutron binding energy exhibit, on the contrary, 
a periphery rich in neutrons - a ``neutron halo'' in the terminology 
introduced 
more than a quarter of a century ago \cite{davis67,nolen69,bugg73b}.

Figure\ \ref{fig5} shows, as a function of the target mass, another 
observable of the 
present experiment, namely the absolute yield (per 1000 antiprotons) of the 
production of nuclei with one nucleon less than the target mass. For all but 
the {\normalsize $^{176}$}Yb target this yield is close to 10\%, without any 
noticeable dependence on the target mass number. The correlation of two 
observables, the halo factor and the yield of (A{\normalsize $_{t}$}-1) 
nuclei, presented in Fig.\ \ref{fig6}, demonstrates again the unusual 
character of the results obtained for {\normalsize $^{176}$}Yb.

\subsection{Isomeric ratios} 

If one of the (A{\normalsize $_{t}$}-1) products has an isomeric state with 
low excitation energy the peripheral antiproton annihilation can populate, 
besides the ground state, also this isomer. Table\ \ref{isomers} shows the 
experimentally determined isomeric ratios. Not considering the 
3$^{-}$ / 0$^{-}$ isomeric pair in $^{152}$Eu (where the experimental limit is
too high to be significant) the high spin / low spin formation ratio is always 
smaller or substantially smaller than one.

This is in marked contrast to the isomeric ratios determined for the deep 
spallation products after stopped antiproton annihilation on nuclei 
\cite{moser89,jastrzebski93b}. In this case, due to the emission of energetic 
particles in the cascade process, the final nuclei acquire a considerable 
angular momentum, what is reflected by a preferential production of high spin 
isomers. In the case of (A{\normalsize $_{t}$} - 1) nuclei the isomeric ratio 
depends on the microscopic composition of the nuclear periphery around the 
annihilation site. The contribution of the high spin components is lowered 
there due to the centrifugal barrier but may be increased due to a larger 
occupation number in comparison with the low spin components. Evidently, 
in the discussion of the isomeric ratio for the (A{\normalsize $_{t}$} - 1) 
nuclei the feeding of isomers by shorter lived states with excitation energies 
below the particle emission threshold should be also considered. Although such 
a discussion for all measured cases would be outside the scope of this 
experimental paper, we present below in some detail one particularly simple 
example ({\normalsize $^{129m,g}$}Te). We hope that other cases given in 
Table\ \ref{isomers} may be sometimes used to furnish a supplementary check 
for the calculations of the nuclear periphery.

\subsection{Charge exchange reactions} 

The charge exchange process was claimed \cite{gerace74} as mainly responsible 
for the larger number of $\pi^{-}$ events in heavy than in light nuclei in 
antiproton annihilation data reported by Bugg et al. \cite{bugg73b}. (Bugg 
attributed this difference to the neutron halo effect in heavy nuclei.) In 
the present method the simulation of a neutron halo or at least the 
increase of the observed effect could be due to the transformation of the 
(N{\normalsize $_{t}$}, Z{\normalsize $_{t}$} - 1) annihilation products to 
(N{\normalsize $_{t}$} - 1, Z{\normalsize $_{t}$}) ones via 
$\pi^{0} \rightarrow \pi^{-}$ or $\pi^{+} \rightarrow \pi^{0}$ charge exchange 
reactions occurring between annihilation pions and inner nucleus (if they 
could proceed without a substantial excitation of this nucleus). Although 
such a transformation is evidently undetectable by our experimental method, 
we can estimate its importance looking for the formation of 
(Z{\normalsize $_{t}$} + 1) nuclei, possible only via the same charge 
exchange processes.

The determined upper limits for the production of  (Z{\normalsize $_{t}$} + 1) 
nuclei are presented in Table\ \ref{limits}. The comparison of the absolute 
yields of 
these nuclei with yields presented in the second and third column of 
Table\ \ref{results} 
clearly indicates that the charge exchange effects are generally much smaller 
than the experimental errors assigned to the production yields of nuclei 
determining the halo factor. Therefore, in agreement with arguments presented 
in Ref. \cite{bugg75}, we will in the present work ignore the corrections 
which could result from these processes. This is not in contradiction with the 
results of Ref. \cite{moser89,egidy90}, where two (Z{\normalsize $_{t}$} + 1) 
products were observed. Their absolute yields were, however, not much larger 
than the upper limit values presented in Table\ \ref{limits}.

\section{DISCUSSION} 

The method presented in this work has allowed us to correlate in 
a quantitative way the neutron enhancement in the nuclear periphery with the 
neutron separation energy values, B{\small $_{n}$}. The data presented in 
Fig.\ \ref{fig3} indicate that isotopes having B{\small $_{n}$} smaller than 
about 10 
MeV exhibit a nuclear periphery in which the neutron density is larger than 
that expected from the N/Z ratio of a given nucleus. As the neutron separation 
energy for the heaviest isotopes of all naturally occurring elements is 
generally below 10 MeV - the enhancement of the nuclear periphery with 
neutrons should be a quite common phenomenon. In the following section we will 
show that this observation is in a qualitative agreement with expectations 
based on the Hartree-Fock-Bogoliubov approach, a model commonly used 
\cite{negele70} to describe the properties of the nuclear periphery. To 
compare the experimental results with theory in a more quantitative way the 
antiproton-nucleus interaction should be considered. This was thoroughly 
discussed in a previous paper by some of us \cite{wycech96} and is also 
briefly outlined below.

\subsection{Nuclear periphery from HFB calculations} 

In the present paper we concentrate on the description of the nuclear 
periphery using a self-consistent Hartree-Fock-Bogoliubov (HFB) model. We are 
perfectly aware of the limitations inherent to this model. One of them, namely 
the unability to correctly predict the binding energies, is probably the most 
troublesome for an approach in which the binding energy is used to correlate 
one of the experimental observables. This was the reason why 
in our previous papers \cite{lubinski94,wycech96} we have also investigated 
another very simple asymptotic density model \cite{bethe70} in which a number 
of phenomenological inputs, including binding energies, was used. However, the 
agreement with the experiment was apparently not improved in comparison with 
the HFB model. Therefore, at the present stage of this research we concentrate 
on the HFB method. We hope that in future works our experimental data will 
allow to discriminate between different approaches \cite{baran97} modelling 
the nuclear periphery. 

We have applied an HFB code which uses the coordinate representation and 
solves the HFB equation on a spatial mesh. For the Skyrme interaction, the HFB 
equation is a differential equation in spatial coordinates. In our 
calculations, we used the SKP version of the Skyrme interaction 
\cite{dobaczewski84}. Due to the imposed spherical symmetry, the HFB+SKP model 
is solved separately for each partial wave ({\it j},{\it l}). We used 100 mesh 
points in the radial coordinate in a box of the size of 25 fm. As a boundary 
condition, we demanded the wave functions to vanish at the far end of the box.

The above described method gives the amplitude of the wave functions which can 
be followed to large distances from the nuclear center. The contribution of 
each orbital to the nuclear density is given by the square of this wave 
function at a given radius times the occupation number for this orbital. 
Particle wave functions are normalized in such a way that the trace of the 
hermitian density matrix for protons (neutrons) is equal to the proton 
(neutron) number. For large radial distances the centrifugal barrier 
suppresses the contribution of high - {\it j} orbitals in comparison with 
those with small - {\it j}. (This effect is illustrated in Fig. 3-4 of Ref. 
\cite{bohr69}). The centrifugal barrier effect may be compensated or overcome 
by the occupation number for high - {\it j} orbitals. As a result, both low- 
and high-spin orbitals contribute to the composition of the nuclear periphery. 
Figure\ \ref{fig7} gives an example of the calculated composition of the 
nuclear 
periphery for the {\normalsize $^{130}$}Te isotope. The antiproton 
annihilation leads to a hole in one of the indicated orbitals. The 
corresponding hole-excitation state will decay by gamma or conversion electron 
emission to the {\normalsize $^{129}$}Te or {\normalsize $^{129}$}Sb ground 
state. In the case of {\normalsize $^{129}$}Te the annihilation can also lead 
to the excitation of the h11/2 isomer. (The 19/2-isomer in 
{\normalsize $^{129}$}Sb has a three-quasi-particle nature \cite{stone87} 
and, evidently, cannot be excited directly by antiproton annihilation 
on any {\normalsize $^{130}$}Te proton orbital. This is in agreement with the 
experimental result given in Table\ \ref{isomers}).

Figure\ \ref{fig7} shows only the amplitudes of orbitals contributing with 
more than 
5\% to the composition of the nuclear periphery. Among states having smaller 
contributions some are so deeply bound that the antiproton annihilation on 
them leads to excitation energies of (A{\normalsize $_{t}$}-1) nuclei larger 
than the neutron separation energy (or proton separation energy + Coulomb 
barrier, or fission barrier). These ``deep hole'' states will evidently decay 
by particle emission, affecting the primary population of the 
(A{\normalsize $_{t}$}-1) nuclei. The total contribution of these states to 
the composition of the nuclear periphery is also indicated in 
Fig.\ \ref{fig7}. Around 
the antiproton annihilation site (discussed in more detail below) this 
contribution does not exceed 30\% for all cases discussed in this work.

Figure\ \ref{fig8} presents the calculated ratio of the (normalized) neutron 
to proton 
densities as a function of the radial distance for some target isotopes 
indicated in Table\ \ref{results}. The experimentally observed neutron rich 
periphery for 
isotopes with neutron separation energy smaller than 10 MeV is qualitatively 
confirmed by the calculated densities. On the contrary, the proton rich 
nuclear atmosphere observed for $^{106}$Cd and $^{144}$Sm 
(f$^{periph}_{halo} < $ 1 is not expected by the theory).

Figure\ \ref{fig9} shows the absolute values of the calculated neutron and 
proton 
density for the {\normalsize $^{96}$}Zr isotope. Figure\ \ref{fig10}, 
finally, displays 
the neutron to proton density ratio as a function of the radial distance for 
a series of even Zr isotopes. The gradual ``build up'' of the neutron 
atmosphere 
with increasing mass number is clearly shown by this calculation. The Figure 
presents, besides stable isotopes the theoretical expectation for isotopes up 
to 8 mass units heavier or lighter than the stable ones. It illustrates what 
may be expected when experiments with moderately changed isospin will become 
feasible.

\subsection{Antiproton-nucleus interaction} 

The interactions in question involve three distinctly separate stages. 1) The 
initial atomic state of the antiproton is determined by the long-range Coulomb 
interaction. At nuclear distances, however, the atomic wave function 
$\Phi_{\bar{N}}$ is determined by the centrifugal barrier and an 
antiproton-nucleus optical potential. 2) The next stage consists of antiproton 
annihilation on a nucleon leading to final mesons. This process lies beyond 
the present theoretical description and must be discussed in a 
phenomenological way in terms of the pion multiplicity and energy 
distribution. 3) The final pions undergo elastic and inelastic nuclear 
reactions. The radiochemical method filters these to the elastic channel.

Each stage is fairly complicated and requires an approximate description. 
Fortunately there are two simplifying effects: the strong absorption of the 
antiproton and pions and the large energy release in the annihilation act. 
It is this energy release that allows to use the closure approximation over 
the final nuclear states. As a consequence one can obtain \cite{wycech96} a 
semi-classical formula for the partial absorption width on a single nucleon 
{\it i,} 
\begin{eqnarray}
\Gamma = 4\frac{\pi}{\mu_{N\bar{N}}}Ima_{N\bar{N}}\int\mid\Phi_{\bar{N}}%
({\bf Y})\mid^{2}\nu({\bf Y}-{\bf X})\rho_{i}({\bf X})(1-P_{dh}({\bf X}))%
P_{miss}(\frac{{\bf X}+{\bf Y}}{2})d{\bf Y}d{\bf X}.\label{one}
\end{eqnarray}

Here {\bf Y} are the antiproton and {\bf X} are the nucleon coordinates, 
respectively. The $\mu_{N\bar{N}}$ is the reduced mass of $N\bar{N}$ system 
and $\rho_{i}$ 
is the nuclear density for protons or neutrons, given by a sum over single 
nucleon states, $\rho = \sum_{a} \phi^{2}_{a}$. This density is folded over 
a form factor $\nu(Y-X)$ that represents the finite extent of the 
annihilation region (radius 0.8 fm). The last two terms describe final state 
effects. $P_{miss}$ is the probability that final state mesons do not interact 
or interact only elastically. This function excludes the annihilation events 
which produce highly excited A-2 or lighter final nuclei. Although 
calculations of $P_{miss}$ are fairly involved the net effect is close to a 
simple geometric estimate of an opening angle as seen by pions to avoid 
nuclear collisions \cite{cugnon97}. Finally $P_{dh}$ is a model correction 
for the rearrangement energy in the final nucleus. It excludes those captures 
that lead to A{\normalsize $_{t}$}-1 nuclei excited above the particle 
emission or fission thresholds.

In the limit $P_{miss}$ = 1, $P_{dh}$ = 0 and $\rho_{i}$ equal to the total 
nuclear density Eq.\ (\ref{one}) produces atomic level widths measured in the 
X-ray 
experiments. $Ima_{N\bar{N}}$ is believed to represent an effective averaged 
antiproton - nucleon absorptive amplitude and is determined by a fit to atomic 
level widths and shifts. This number is not needed for the studies of neutron 
halo. What is needed is the ratio $R_{np} = Ima_{n\bar{p}}/Ima_{p\bar{p}}$ 
which, as indicated above, is taken as 0.63. (In Ref. \cite{wycech96} the 
arguments to use the slightly different value R{\normalsize $_{np}$}=0.82 were 
given. In the present paper we keep our previous \cite{lubinski94} definition 
of the halo factor in order to facilitate the comparisons.) For the details of 
final state calculations and the derivation of the intuitively simple formula 
(1), we refer to Ref. \cite{wycech96}. One problem in the analysis of halo 
factors is to know the atomic state from which nuclear capture takes place. At 
this stage of research a complete answer is not available. One can calculate 
the distribution of these states for the lower part of the atomic cascade. 
This is possible for states {\it n}$ < $20, i.e. states localized between 
nucleus and electron cloud. In addition, this part of the cascade may be 
checked directly via measurements of X-ray intensities. It was concluded 
\cite{wycech96,wycech95} that about 95\% of the nuclear captures take place 
from levels with only one or two different values of the antiproton angular 
momentum {\it  l}, although the distribution over {\it n} may involve more 
states. These {\it l} values correspond to the circular orbits of the so 
called ``upper'' and ``lower''states of capture. (In the hadronic atom 
terminology the ``lower {\it n}'' state is the last state, which can be 
observed before the strong interaction prevents further hadron cascading and
X-ray emission. The ``upper'' level has the principal quantum number {\it n} 
higher by one in respect to the ``lower'' level.) The difficulty is that 
these, low {\it n}, captures come to 
at most half of the total nuclear capture rate \cite{wiegand74,trzcinska98}. 
In our calculations the {\it l} distribution of capture states given by the 
low {\it n} cascade calculations have been used.

The arguments in favour of this assumption are illustrated in 
Fig.\ \ref{fig11}. In this 
Figure the calculated total absorption probability densities in antiprotonic 
Nd and the corresponding distributions for cold absorption (i.e. those, 
leading to the production of A{\normalsize $_{t}$}-1 nuclei) are shown for 
several states with different antiproton angular momentum. These densities 
are, for a given value of {\it l}, almost {\it n}-independent. In the Nd atom 
the ``low {\it n}'' capture takes place from the {\it l}=6 and {\it l}=7 
states. 
The overlap of the antiproton wave function with the nucleus in these and 
higher {\it l} states is localized at the nuclear surface by the centrifugal 
barrier. On the other hand, for lower values of {\it l} another localizing 
factor arises - the strong absorption of antiprotons. As a result, the radius 
of maximum absorption reaches some limiting value. This penetration blocking 
indicates that even if some of the antiprotons are absorbed from lower {\it l} 
states the spatial scenario of cold capture is not changed significantly.

\subsection{Comparison of the experimental data with calculations} 

\subsubsection{Annihilation site} 

As it was indicated at the beginning of this paper, the experiment determines 
the neutron to proton density ratio at the nuclear periphery, presumably at 
distances close to the antiproton annihilation site. The calculations 
mentioned above and presented in detail in Ref. \cite{wycech96} indicate that, 
almost independently of the target mass, the method is most sensitive for 
densities encountered at distances about 3 fm larger than the half-density 
radius. The width of the annihilation probability distribution (FWHM) for 
events testing the density ratio is between 2 and 3 fm (compare also 
Fig.\ \ref{fig11}).

The experimental observable related to the annihilation site is the production 
yield of nuclei one mass unit lighter than the target mass. This yield, 
related to the annihilation geometry, depends strongly on the antiproton - 
nucleus interaction, pion - nucleus interaction, pion energy distribution and 
kinematic correlations. Fortunately, it is rather insensitive to the neutron 
to proton density ratio, which affects mainly the second observable of this 
experiment.

The rather good agreement between the experimental data and theoretical 
estimate for the production yield of A{\normalsize $_{t}$}-1 nuclei, presented 
in Fig.\ \ref{fig12}, indicates that all factors governing this production 
are to a 
large extent understood. However, the only large exception, the 
{\normalsize $^{176}$}Yb nucleus, for which the annihilation site seems to be 
located much farther away from the nucleus than predicted by calculations 
needs further studies (cf. \cite{wycech96}).

\subsubsection{Neutron to proton density ratio} 

Figure\ \ref{fig13} shows a comparison between the experimentally determined 
and 
calculated  values of the halo factor. Again, the agreement is fair although 
slightly less good than for the yield of A{\normalsize $_{t}$}-1 products, 
indicating the need for some improvements in the nuclear density calculations. 
This is especially true for the two ``proton halo'' nuclei 
{\normalsize $^{106}$}Cd and {\normalsize $^{144}$}Sm for which the HFB 
calculations predict a neutron rather than a proton rich atmosphere at large 
nuclear distances (see Fig.\ \ref{fig8}). On the contrary, the calculated 
neutron and 
proton densities in {\normalsize $^{176}$}Yb can be reasonably close to 
reality, the discrepancy between theory and experiment lying mainly in the 
underestimated annihilation distance.

\subsubsection{Microscopic composition of the nuclear periphery from the 
isomeric ratio}

Lets us discuss, as an example, the case of the {\normalsize $^{129}$}Te 
isomeric ratio. We assume that the decay of the excited hole states in 
{\normalsize $^{129}$}Te after the antiproton annihilation on the peripheral 
neutron proceeds without the parity change (E1 transitions much weaker than M1 
and E2 ones). From calculations for the {\small $^{130}$}Te target, as 
indicated above, one deduces that the annihilation site is located around r = 
8.7 fm in case of the antiproton-neutron interaction. At this distance the 
ratio of the 1h11/2 orbital density (main component of the high spin isomer) 
to the sum of densities of the 2d3/2 and other positive parity orbitals is 
0.26. As no other negative parity orbitals contribute to the nuclear density 
at this distance this is the approximate theoretical expectation of the 
isomeric ratio, experimentally determined (cf. Table\ \ref{isomers}) to be 
0.45$\pm$0.15. 
In a more refined approach one would consider the antiproton annihilation 
separately for each orbital (similarly, as it was discussed in \cite{bloom72}) 
and calculate the sum of the corresponding negative and positive parity 
contributions. The assumption of the spectroscopic factors S=1 should not 
introduce any appreciable errors, as states belonging to the same orbital 
should be interconnected by fast gamma transitions.

Similar arguments applied to the {\normalsize $^{95}$}Tc isomeric ratio lead 
to a theoretical value equal to 0.28, to be compared with the experimental 
result of 0.58$\pm$0.22.

\section{SUMMARY AND CONCLUSIONS} 

Using the recently proposed new method for the study of the nuclear periphery 
composition we have investigated a number of targets in the mass range from 58 
to 238. The experimental data clearly indicate that the periphery of nuclei in 
which the neutron binding energy is smaller than about 10 MeV has more 
neutrons than would be expected from the target N/Z ratio. As almost all heavy 
stable isotopes of all elements exhibit neutron binding energy smaller than 
the value quoted above - the neutron rich periphery should be a quite common 
phenomenon for stable and even more for neutron rich radioactive isotopes. 
However, the observation of this phenomenon may be difficult in experiments 
in which the nuclear periphery is tested with methods less sensitive than 
offered by strongly interacting probes. Indeed, the calculations based on the 
antiproton - nucleus and pion - nucleus interactions indicate that the method 
used in the present work is sensitive at nuclear distances around 3 fm larger 
than the nuclear half-density radius. At these distances the nuclear density 
is about 100 to 1000 times smaller than the central one, which leads to the 
evident experimental difficulties.

In two cases investigated in this work it was found that the outer periphery 
of the nuclei studied ({\normalsize $^{106}$}Cd and {\normalsize $^{144}$}Sm) 
exhibit a proton rich atmosphere. Although this result may be understood 
intuitively at least for the closed neutron shell nucleus 
{\normalsize $^{144}$}Sm, it contradicts the nuclear periphery composition 
expected from the Hartree-Fock-Bogoliubov calculation performed in this work. 
Therefore it can probably be used as a sensitive test for the nuclear 
periphery models.

The experimentally determined isomeric ratio for nuclei one mass unit lighter 
than the target mass will probably constitute a similarly sensitive test for 
modelling the nuclear periphery. If this happens to be true, a number of other 
cases could be investigated in the near future using the new antiproton 
facility at CERN.

\acknowledgments

Our thanks are due to Peter Maier-Komor, Anna Stolarz and Katharina Nacke for 
their invaluable help in preparing the targets and to Anna Grochulska, who 
participated at the early stage of this work. We also gratefully acknowledge 
the fruitful discussions with Jacek Dobaczewski and Janusz Skalski. The 
competent and efficient help of the LEAR staff during the irradiations was 
crucial for the success of this work. 

The present paper summarises the results of many years of experiments and 
evaluations which were continuously supported by the research and travel 
grants from the Polish State Committee for the Scientific Research, by the 
Joint Project of Science and Technology Cooperation between Germany and Poland 
and recently by the grant from the Volkswagen Foundation.

\begin{figure}
\caption{Example of the measured antiproton range distribution in an Al stack, 
determined by counting the produced {\normalsize $^{24}$}Na activity. Before 
impinging on the Al stack the antiproton energy, equal to 21 MeV, was degraded 
in 800 $\mu$m of mylar and in 2.5 mm of a plastic scintillator.} 
\label{fig1}
\end{figure}

\begin{figure}
\caption{Mass yield distribution for the $^{130}$Te target. The method for 
the determination of this distribution is given elsewhere [40,58]. }
\label{fig2}
\end{figure}

\begin{figure}
\caption{Neutron halo factor (defined in the text) as a function of the target 
neutron separation energy, B{\small $_{n}$}.}
\label{fig3}
\end{figure}

\begin{figure}
\caption{Absolute production yield of isotopes A{\small $_{t}$}-1 having one 
mass unit less than the target mass as a function of the target mass number.}
\label{fig5}
\end{figure}

\begin{figure}
\caption{Correlation between halo factor and absolute production yield for 
A{\small $_{t}$}-1 nuclei.}
\label{fig6}
\end{figure}

\begin{figure}
\caption{Relative contribution of the individual shell model orbitals 
to the total neutron and proton densities in {\normalsize $^{130}$}Te as 
a function of the radial distance, obtained using the Hartree-Fock-Bogoliubov 
approach. The curve marked P{\normalsize $_{dh}$} shows the contribution of 
orbitals bound by more than the neutron separation 
energy. The arrows indicate the most probable distance of annihilation 
on a neutron (upper part) and on a proton (lower part), respectively.}
\label{fig7}
\end{figure}

\begin{figure}
\caption{Calculated (HFB method) neutron to proton density ratio as a function 
of the nuclear distance for some investigated isotopes.}
\label{fig8}
\end{figure}

\begin{figure}
\caption{Calculated neutron and proton densities for {\normalsize $^{96}$}Zr.}
\label{fig9}
\end{figure}

\begin{figure}
\caption{The same as Fig. 7 for Zr isotopes. The dashed lines show the density 
ratio for the unstable Zr isotopes.}
\label{fig10}
\end{figure}

\begin{figure}
\caption{Total probability densities (full lines) for antiproton absorption 
from circular orbits in Nd as a function of the distance from the nuclear 
center, together with the corresponding cold-absorption densities (dashed 
lines). For these curves the scale labelled W({\it l}) holds. The numbers at 
the curves indicate the respective values of the orbital angular momentum. 
The total nuclear density $\rho$(r) (left hand scale) is also shown.}
\label{fig11}
\end{figure}

\begin{figure}
\caption{Comparison of the experimentally determined yield of 
(A{\normalsize $_{t}$}-1) nuclei with the calculated values.}
\label{fig12}
\end{figure}

\begin{figure}
\caption{Comparison of the experimentally determined halo factor 
with the calculated values.}
\label{fig13}
\end{figure}

\newpage

\mediumtext

\begin{table}
\caption{Characteristics of the antiproton beam.}
\label{ranges}
\begin{tabular}{cddccd}
Experiment&\multicolumn{1}{c}{Beam}&\multicolumn{1}{c}{Beam}&
\multicolumn{2}{c}{Measured pathlenght straggling}&
\multicolumn{1}{c}{Calculated pathlenght} \\
&momentum&energy&Target&FWHM&straggling\tablenotemark[1] \\
&[MeV/c]&[MeV]&&[mg/cm$^{2}$]&FWHM [mg/cm$^{2}$] \\
\tableline
1991 & 205.5 & 22.5 & $^{232}$Th & 64$\pm$4 & 57.4 \\
& & & $^{197}$Au & 56$\pm$3 & 57.4 \\
1992 & 200.4 & 21.4 & $^{27}$Al & 22.5$\pm$1.0 & 21.9 \\
1993 & 200.4 & 21.4 & $^{27}$Al & 21.9$\pm$1.0 & 21.9 \\
1995 & 310.1 & 49.9 &&& \\
1996 & 106.0 & 6.0 & $^{27}$Al & 4.6$\pm$0.4 & 2.9 \\
\end{tabular}
\tablenotetext[1]{Ref. \ \cite{janni82}.}
\end{table}

\begin{table}
\caption{Targets and numbers of stopped antiprotons.}
\label{targets}
\begin{tabular}{cccdcc}
Target&Enrichment&Physical form&Thickness&$\bar{p}$ stopped&Experiment \\
 & [\%] & & [mg/cm$^{2}$] & in target [10$^{8}$] & \\
\tableline
$^{45}$Sc & 99.9 & metal & 55.8 & 3.4$\pm$0.2 & 1992 \\
$^{56}$Fe & 99.9 & metal & 43.6 & 9.4$\pm$0.9 & 1993 \\
$^{58}$Ni & 99.9 & metal & 40.1 & 7.4$\pm$0.3 & 1992 \\
$^{58}$Ni & 99.8 & metal & 60.1 & 2.9$\pm$0.2 & 1996 \\
$^{96}$Zr & 85.3 & metal & 35.2 & 2.2$\pm$0.3 & 1992 \\
$^{96}$Zr & 85.3 & metal & 35.2 & 5.6$\pm$0.3 & 1996 \\
$^{96}$Ru & 97.9 & powder + glue & 20.1 & 0.7$\pm$0.2 & 1992 \\
$^{96}$Ru & 97.9 & powder + glue & 20.1 & 0.7$\pm$0.2 & 1996 \\
$^{nat}$Cd & $^{106}$Cd (1.26\%) & metal & 37.5 & 1.8$\pm$0.2 & 1993 \\
$^{nat}$Cd & $^{106}$Cd (1.26\%) & metal & 39.0 & 7.1$\pm$0.3 & 1996 \\
$^{106}$Cd & 76.5 & metal & 40.0 & 2.1$\pm$0.3 & 1993 \\
$^{106}$Cd & 76.5 & metal & 40.0 & 7.6$\pm$0.4 & 1996 \\
$^{128}$Te & 98.3 & metal & 353 & 9.1$\pm$1.0 & 1995 \\
$^{128}$Te & 98.3 & metal & 91.1 & 3.2$\pm$0.2 & 1996 \\
$^{nat}$Te & $^{130}$Te (33.9\%) & powder + glue & 104 & 0.9$\pm$0.2 & 1992 \\
$^{130}$Te & 99.6 & powder + glue & 37.8 & 9.3$\pm$2.0 & 1993 \\
$^{130}$Te & 99.3 & powder + glue & 93.6 & 2.3$\pm$0.2 & 1996 \\
$^{144}$Sm & 96.5 & metal & 47.2 & 3.1$\pm$0.2 & 1992 \\
$^{144}$Sm & 86.6 & metal & 45.7 & 14$\pm$3 & 1993 \\
$^{144}$Sm & 90.8 & metal & 109 & 6.1$\pm$0.3 & 1996 \\
$^{148}$Nd & 91.6 & powder, oxide & 95 & 1.8$\pm$0.4 & 1993 \\
$^{148}$Nd & 88.6 & powder, oxide & 90 & 4.1$\pm$0.4 & 1996 \\
$^{nat}$Eu & - & powder + glue, oxide & 105 & 3.2$\pm$0.4 & 1992 \\
$^{154}$Sm & 97.7 & metal & 54.4 & 3.2$\pm$0.5 & 1992 \\
$^{154}$Sm & 98.3 & metal & 107 & 1.3$\pm$0.1 & 1996 \\
$^{160}$Gd & 98.1 & powder + glue, oxide & 40.7 & 9.8$\pm$2.0 & 1993 \\
$^{160}$Gd & 98.1 & powder + glue, oxide & 110 & 4.4$\pm$0.4 & 1996 \\
$^{nat}$Yb & $^{176}$Yb (12.7\%) & metal & 102 & 0.8$\pm$0.1 & 1992 \\
$^{176}$Yb & 96.4 & metal & 31.1 & 9.8$\pm$2.0 & 1993 \\
$^{176}$Yb & 96.4 & metal & 114 & 8.6$\pm$0.5 & 1996 \\
$^{206}$Pb & 95.9 & metal & 85.7 & 15$\pm$2 & 1993 \\
$^{232}$Th & 100 & metal & 100 & 23$\pm$1 & 1991 \\
$^{232}$Th & 100 & metal & 4.54 & 0.9$\pm$0.3 & 1993 \\
$^{232}$Th & 100 & metal & 43.0 & 1.5$\pm$0.3 & 1993 \\
$^{238}$U & 99.8 & metal & 107 & 5.2$\pm$0.2 & 1992 \\
\end{tabular}
\end{table}

\begin{table}
\caption{Absolute production yield of A$_{t}$-1 nuclei, their yield ratio, and 
peripheral halo factor.}
\label{results}
\begin{tabular}{cccccc}
&\multicolumn{3}{c}{Produced nuclei}&& \\
Target&\large $\frac{N_{t}-1}{1000\bar{p}}$&
\large $\frac{Z_{t}-1}{1000\bar{p}}$&
\large $\frac{A_{t}-1}{1000\bar{p}}$&
\large $\frac{N(\bar{p}n)}{N(\bar{p}p)}$&
f$^{periph}_{halo}$ \\
\tableline
$^{58}_{28}$Ni&45$\pm$4&49$\pm$7&94$\pm$6&0.90$\pm$0.12&1.3$\pm$0.2 \\
$^{96}_{40}$Zr&111$\pm$18&34$\pm$8&145$\pm$11&3.3$\pm$0.6&3.7$\pm$0.6 \\
$^{96}_{44}$Ru&39$\pm$10&50$\pm$14&89$\pm$16&0.79$\pm$0.17&1.1$\pm$0.2 \\
$^{106}_{48}$Cd&33$\pm$6&72$\pm$11&105$\pm$8&0.5$\pm$0.1&0.6$\pm$0.1 \\
$^{128}_{52}$Te&65$\pm$17&17$\pm$3&82$\pm$14&3.9$\pm$1.0&4.3$\pm$1.1 \\
$^{130}_{52}$Te&81$\pm$16&20$\pm$4&101$\pm$12&4.0$\pm$0.4&4.2$\pm$0.4 \\
$^{144}_{62}$Sm&$\leq$ 31&94$\pm$20&110$\pm$8&$\leq$ 0.4&$\leq$ 0.5 \\
$^{148}_{60}$Nd&56$\pm$7&13$\pm$4&69$\pm$8&4.4$\pm$0.9&4.8$\pm$0.9 \\
$^{154}_{62}$Sm&77$\pm$11&39$\pm$8&116$\pm$10&2.0$\pm$0.3&2.2$\pm$0.4 \\
$^{160}_{64}$Gd&94$\pm$23&17$\pm$5&111$\pm$25&5.5$\pm$1.8&5.8$\pm$1.9 \\
$^{176}_{70}$Yb&196$\pm$28&26$\pm$6&222$\pm$25&7.6$\pm$0.6&8.0$\pm$0.6 \\
$^{232}_{90}$Th&71$\pm$12&13$\pm$2&84$\pm$11&5.4$\pm$0.8&5.4$\pm$0.8 \\
$^{238}_{92}$U&91$\pm$7&19$\pm$2&110$\pm$7&5.8$\pm$0.8&5.8$\pm$0.8 \\
\end{tabular}
\end{table}

\narrowtext
\begin{table}
\caption{Isomeric ratios for A$_{t}$-1 nuclei.}
\label{isomers}
\begin{tabular}{cccc}
Isomer & Spin & Energy [keV] & Isomeric ratio \\
\tableline
$^{44}$Sc(m)/$^{44}$Sc(g) & 6$^{+}$/2$^{+}$ & 271/0 & 0.42$\pm$0.05 \\
$^{95}$Tc(g)/$^{95}$Tc(m) & $\frac{9}{2}^{+}$/$\frac{1}{2}^{-}$ & 0/39 & 
0.58$\pm$0.22 \\
$^{127}$Te(m)/$^{127}$Te(g) & $\frac{11}{2}^{-}$/$\frac{3}{2}^{+}$ & 88/0 & 
$\leq$ 0.6 \\
$^{129}$Te(m)/$^{129}$Te(g) & $\frac{11}{2}^{-}$/$\frac{3}{2}^{+}$ & 105/0 & 
0.45$\pm$0.15 \\
$^{129}$Sb(m)/$^{129}$Sb(g) & $\frac{19}{2}^{-}$/$\frac{7}{2}^{+}$ & 1851/0 & 
$\leq$ 0.02 \\
$^{150}$Eu(m)/$^{150}$Eu(g) & 5$^{(-)}$/0$^{(-)}$ & 42/0 & $\leq$ 1.3 \\
$^{152}$Eu(2)/$^{152}$Eu(g) & (8)$^{-}$/3$^{-}$ & 148/0 & $\leq$ 0.016 \\
$^{152}$Eu(g)/$^{152}$Eu(1) & 3$^{-}$/0$^{-}$ & 0/46 & $\leq$ 5.7 \\
$^{152}$Eu(2)/$^{152}$Eu(1) & (8)$^{-}$/0$^{-}$ & 148/46 & 0.11$\pm$0.03 \\
$^{196}$Au(m)/$^{196}$Au(g) & 12$^{-}$/2$^{-}$ & 595/0 & $\leq$ 0.02 \\
\end{tabular}
\end{table}

\begin{table}
\caption{Upper limits for the production of Z$_{t}$+1 isotopes by pion charge 
exchange.}
\label{limits}
\begin{tabular}{ccccc}
Target & Isotope & Spin & \large $\frac{N(Z_{t}+1)}{1000\bar{p})}$ &
\large $\frac{N(A,Z_{t}+1)}{N(A_{t},Z_{t})}$ \\
\tableline
$^{56}$Fe & $^{55}$Co & 7/2$^{-}$ & $\leq$ 0.24 &
$\leq$ 0.006\tablenotemark[1] \\
$^{96}$Zr & $^{95}$Nb(m) & 1/2$^{-}$ & $\leq$ 1.3 & $\leq$ 0.038 \\
& $^{95}$Nb(g) & 9/2$^{+}$ & $\leq$ 0.5 & $\leq$ 0.015 \\
$^{96}$Ru & $^{95}$Rh(m) & 1/2$^{-}$ & $\leq$ 6.2 & $\leq$ 0.12 \\
& $^{95}$Rh(g) & 5/2$^{+}$ & $\leq$ 5.5 & $\leq$ 0.11 \\
$^{206}$Pb & $^{205}$Bi & 9/2$^{-}$ & $\leq$ 0.7 &
$\leq$ 0.009\tablenotemark[2] \\
& $^{204}$Bi & 6$^{+}$ & $\leq$ 0.7 &
$\leq$ 0.009\tablenotemark[2] \\
& $^{203}$Bi & 9/2$^{-}$ & $\leq$ 1.8 &
$\leq$ 0.023\tablenotemark[2] \\
\end{tabular}
\tablenotetext[1]{Production of $Z_{t}$ nuclei assumed to be equal to 40 per 
1000 antiprotons, from systematics based on N($Z_{t}$) dependence on the 
neutron binding energy, B$_{n}$.}
\tablenotetext[2]{Production of $Z_{t}$ nuclei assumed to be equal to 80 per
1000 antiprotons.}
\end{table}

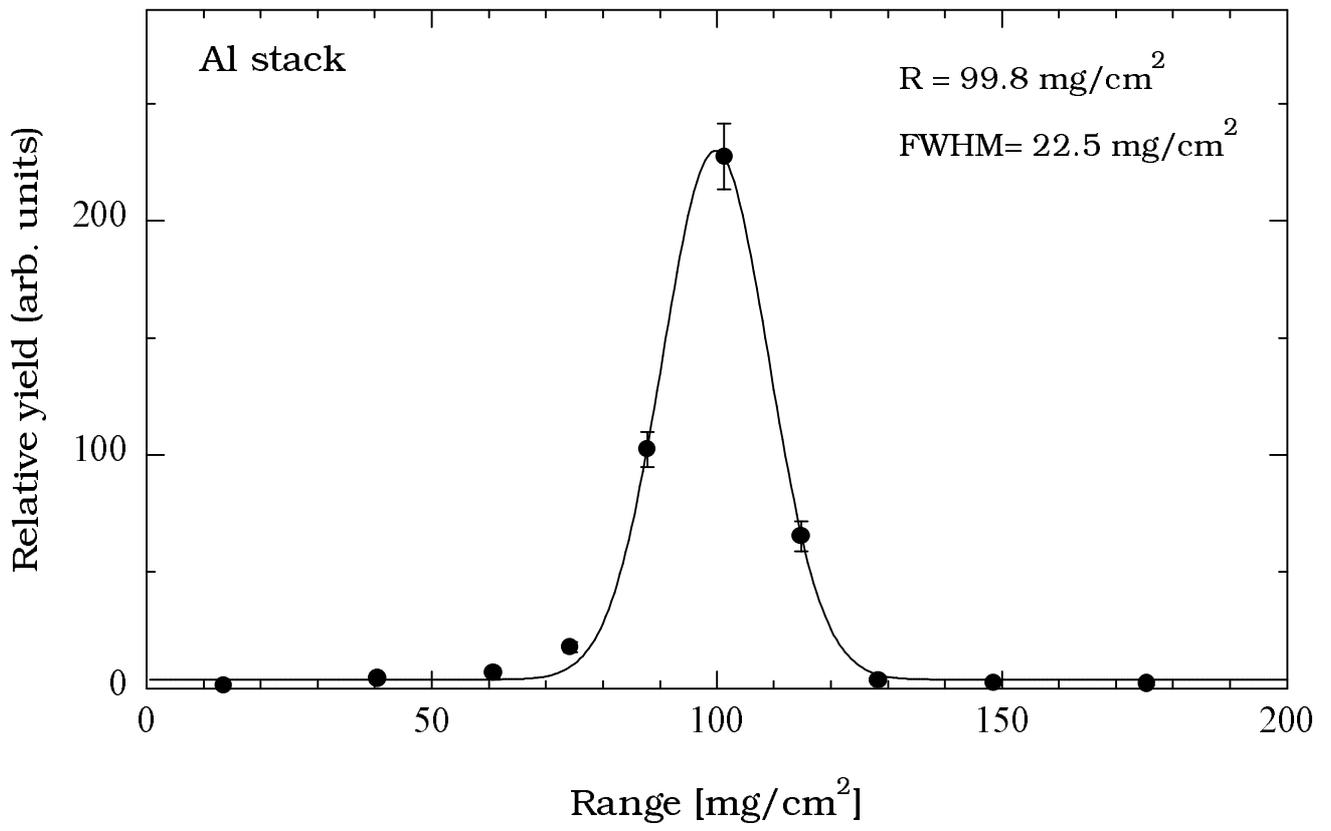
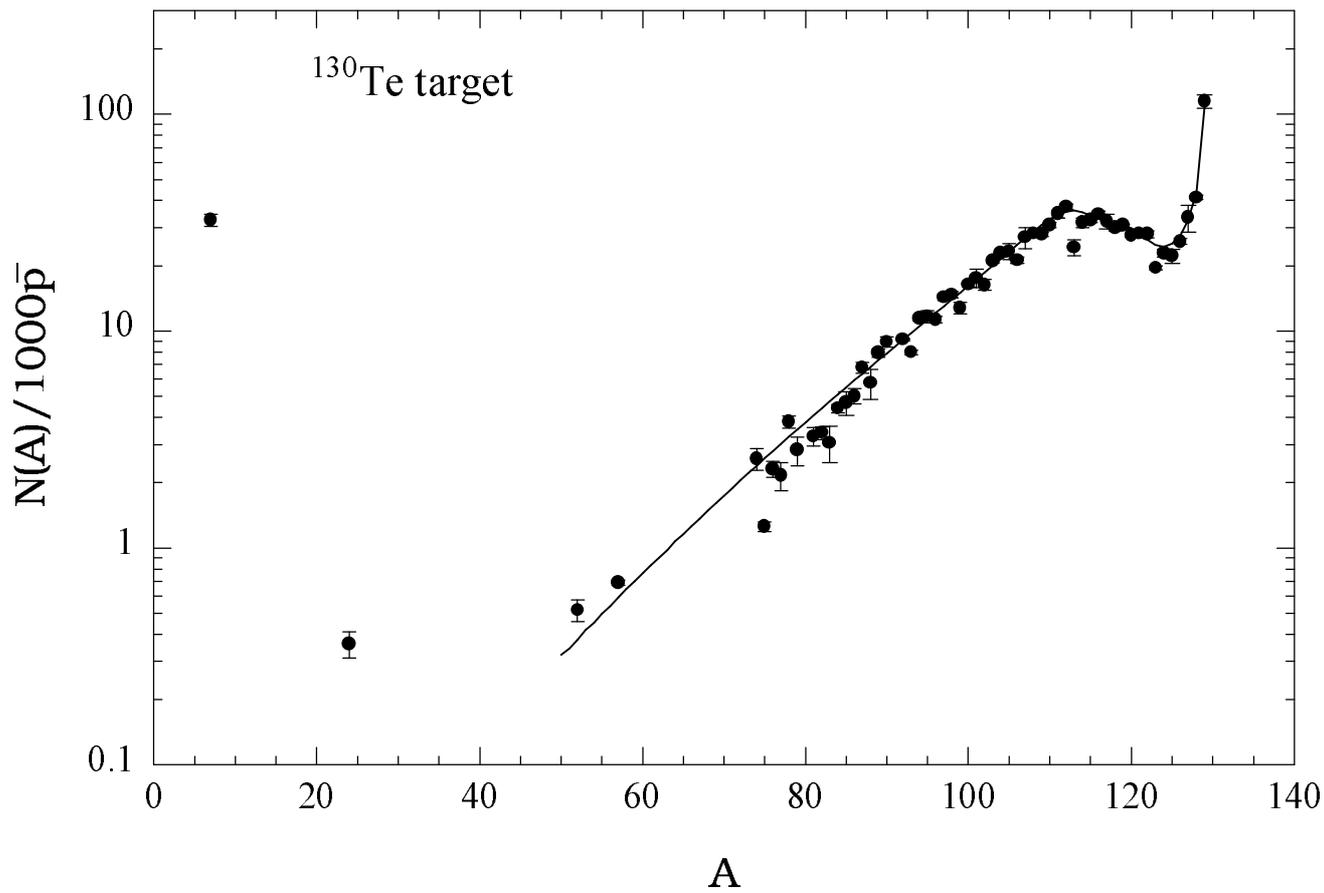
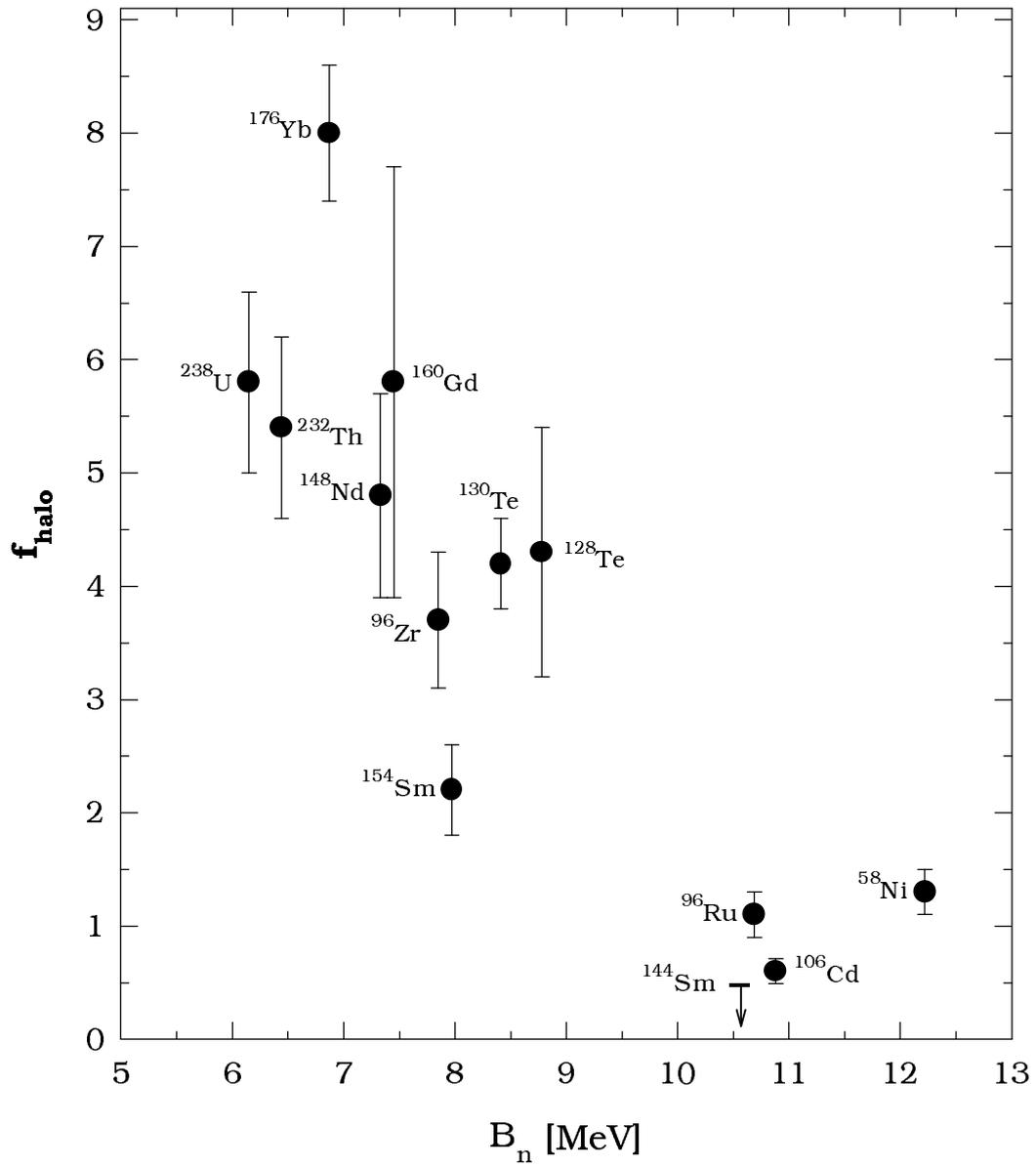
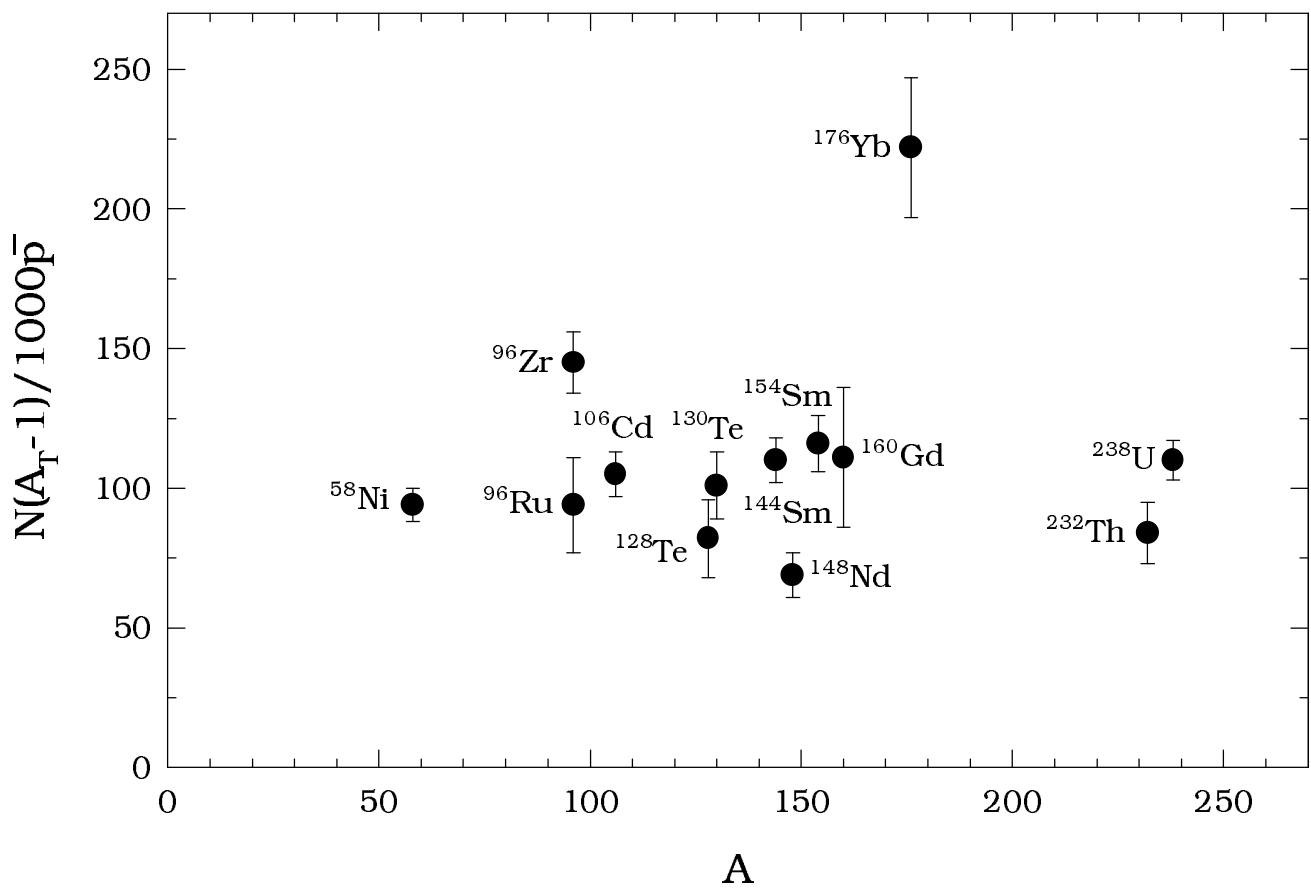
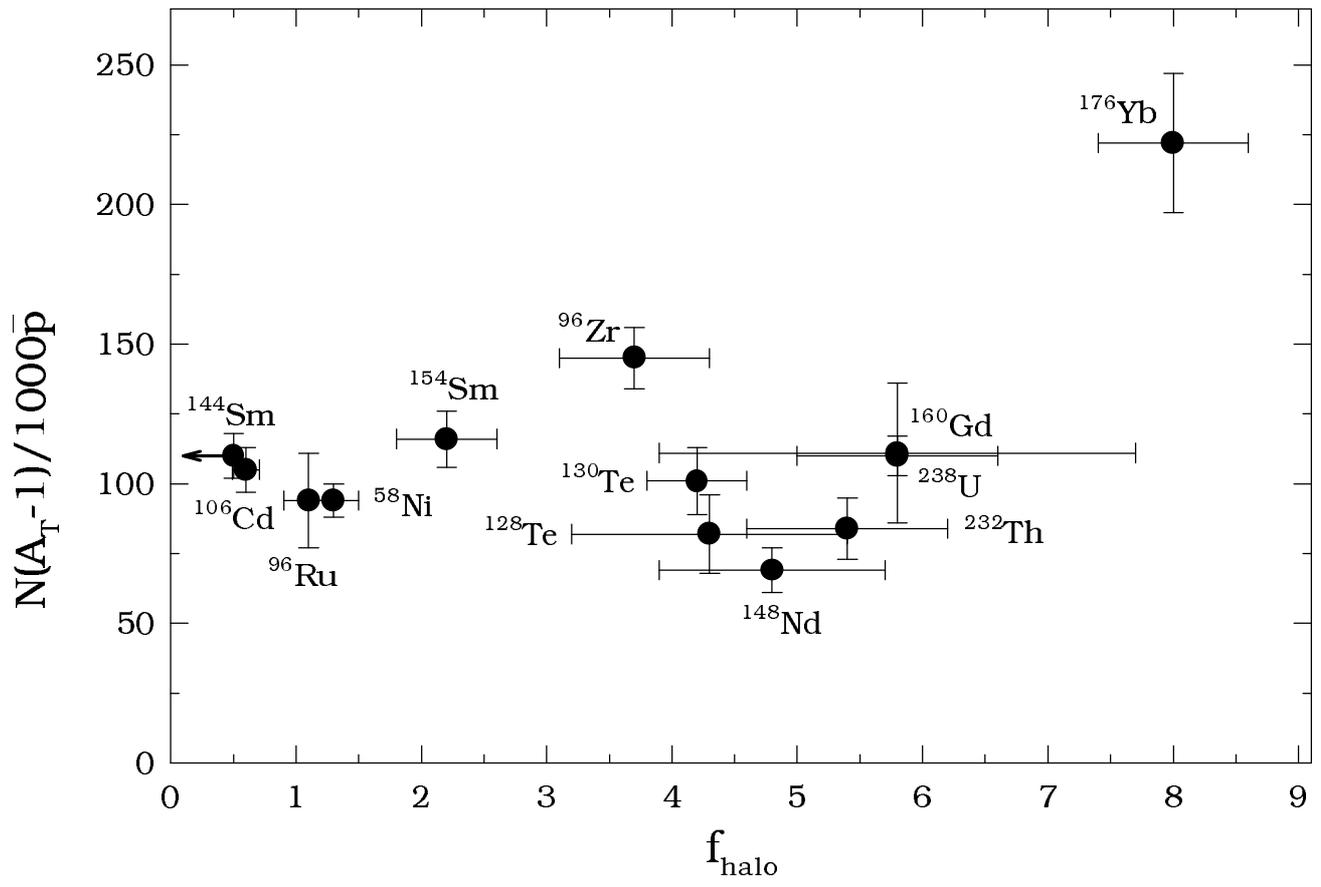
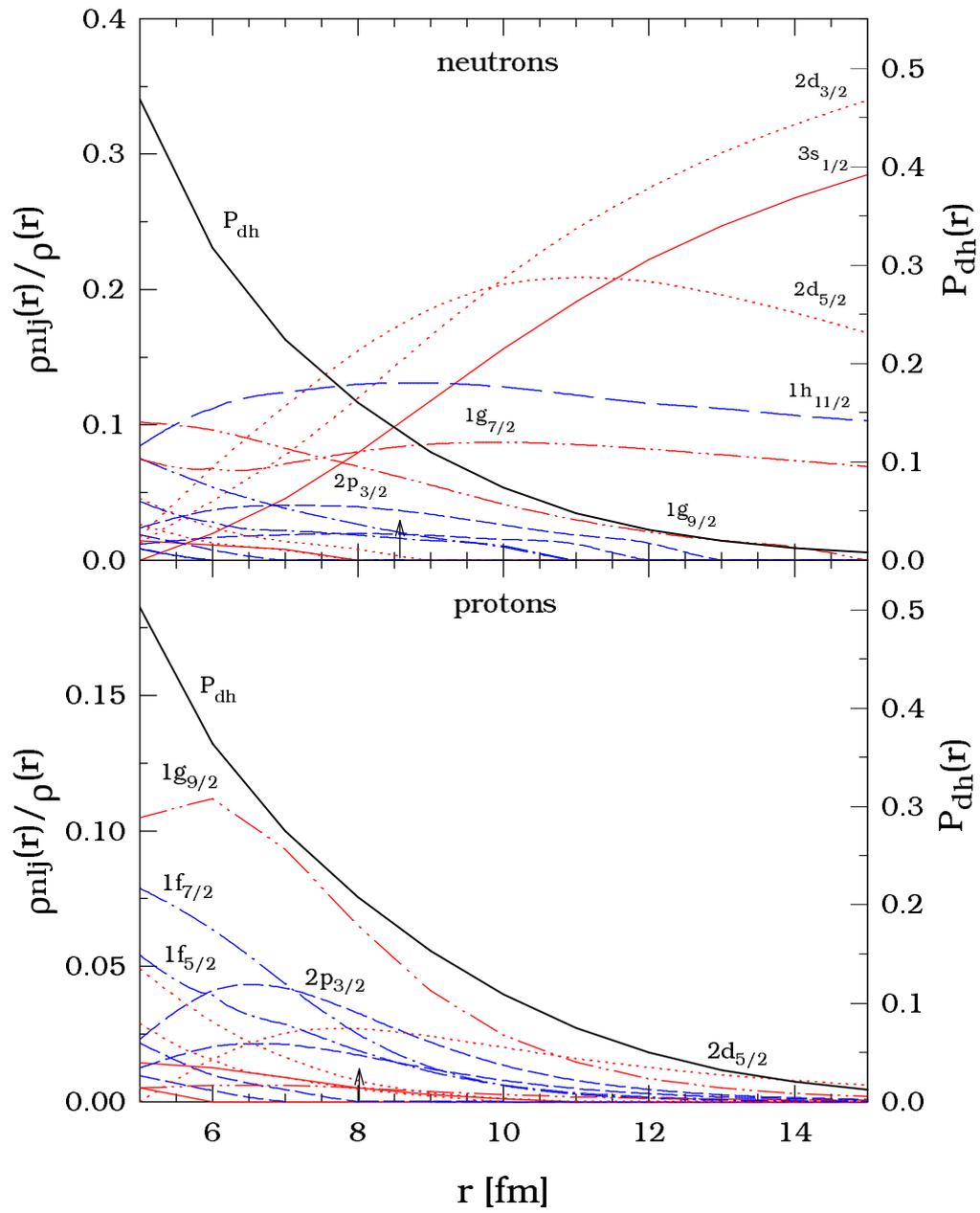
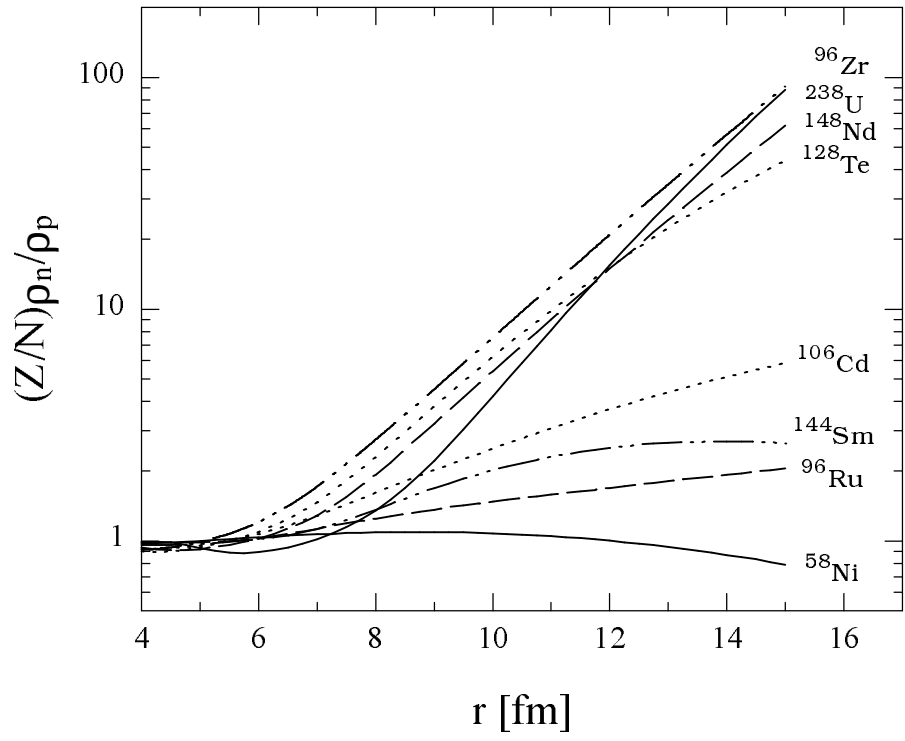
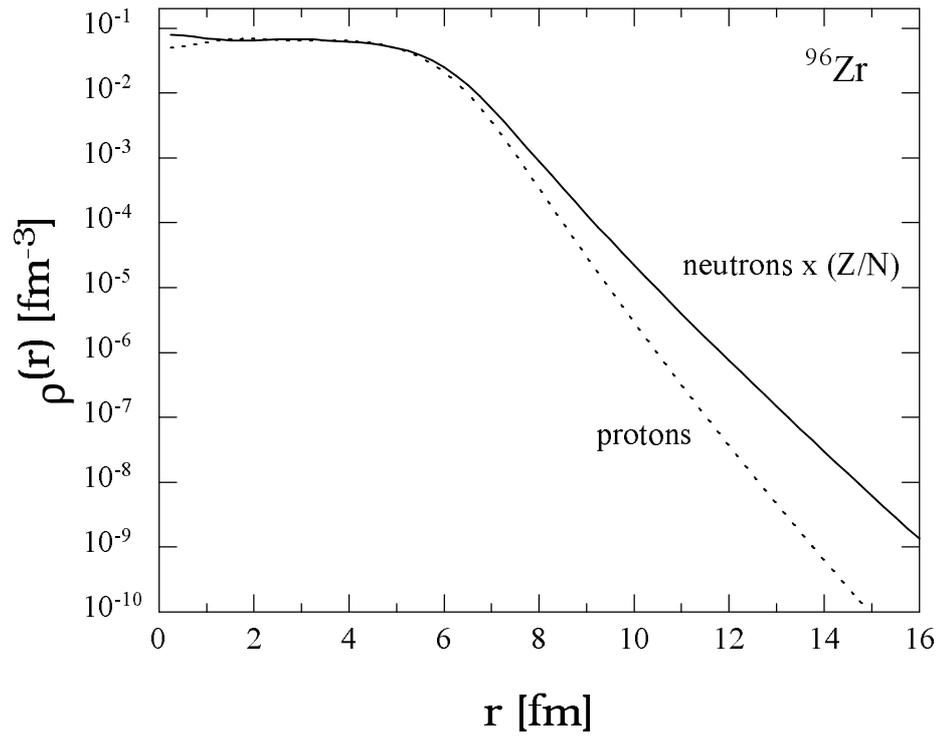
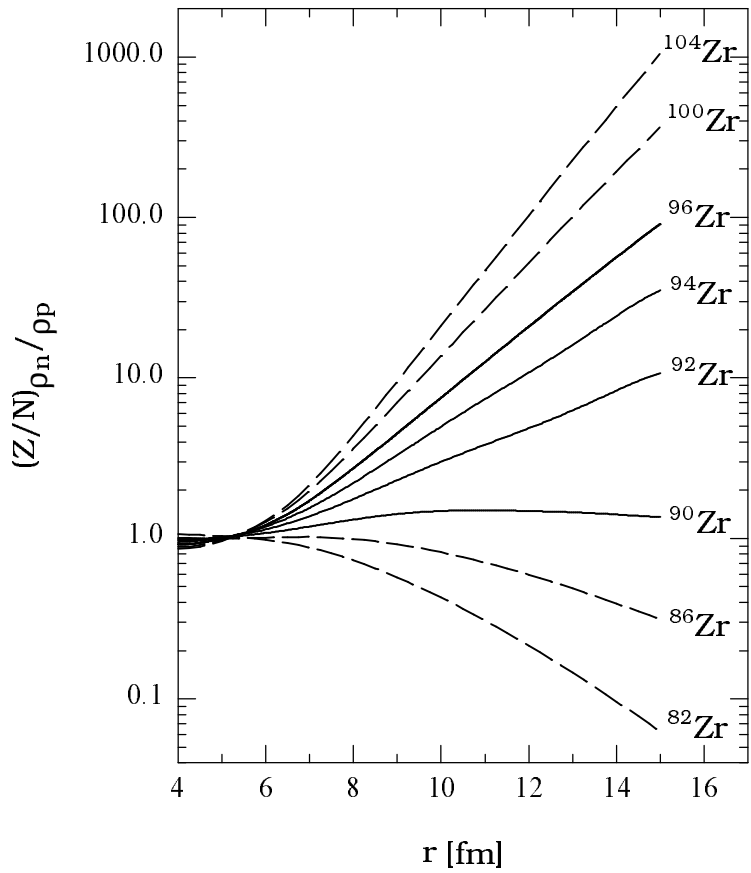
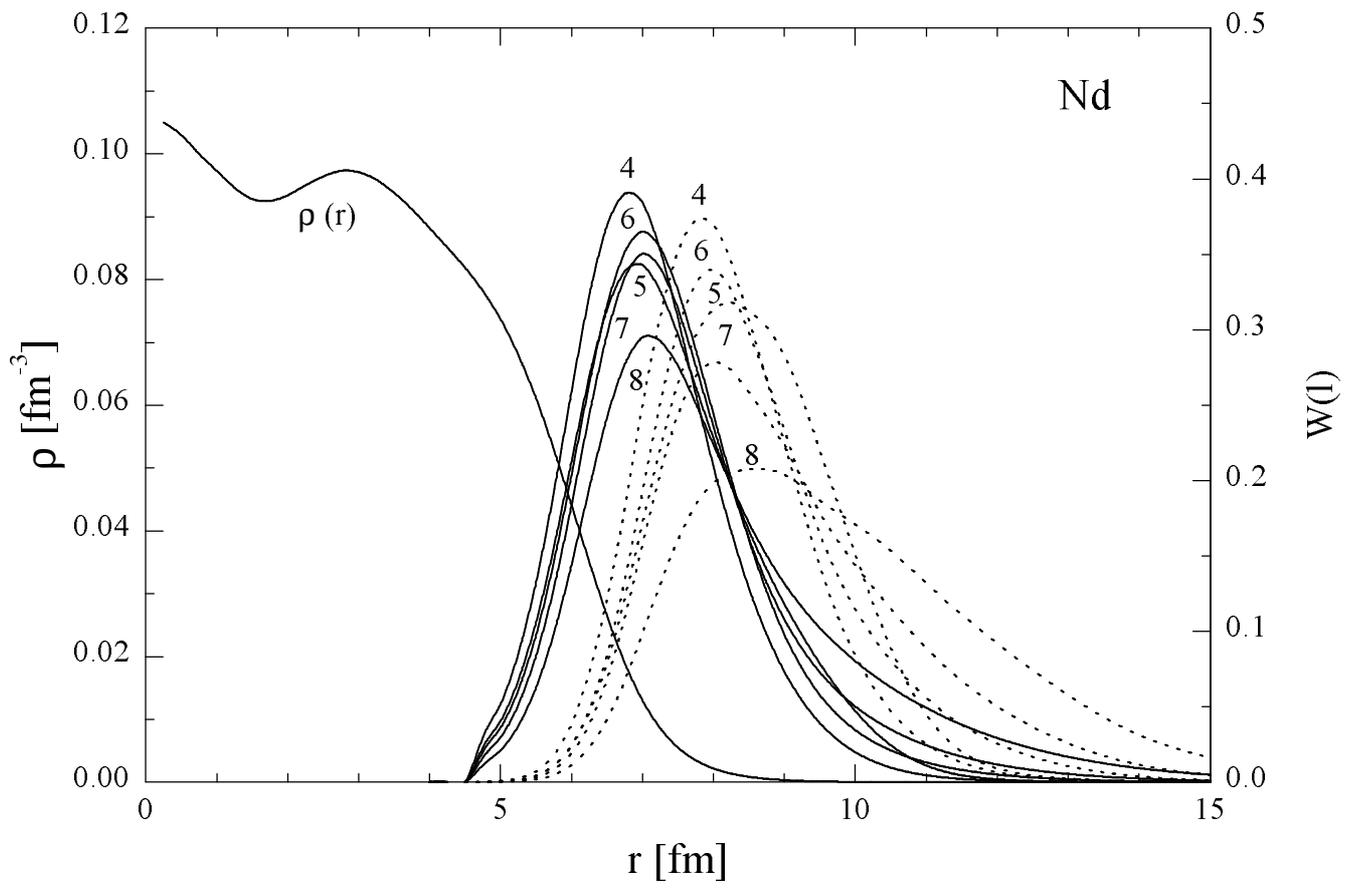
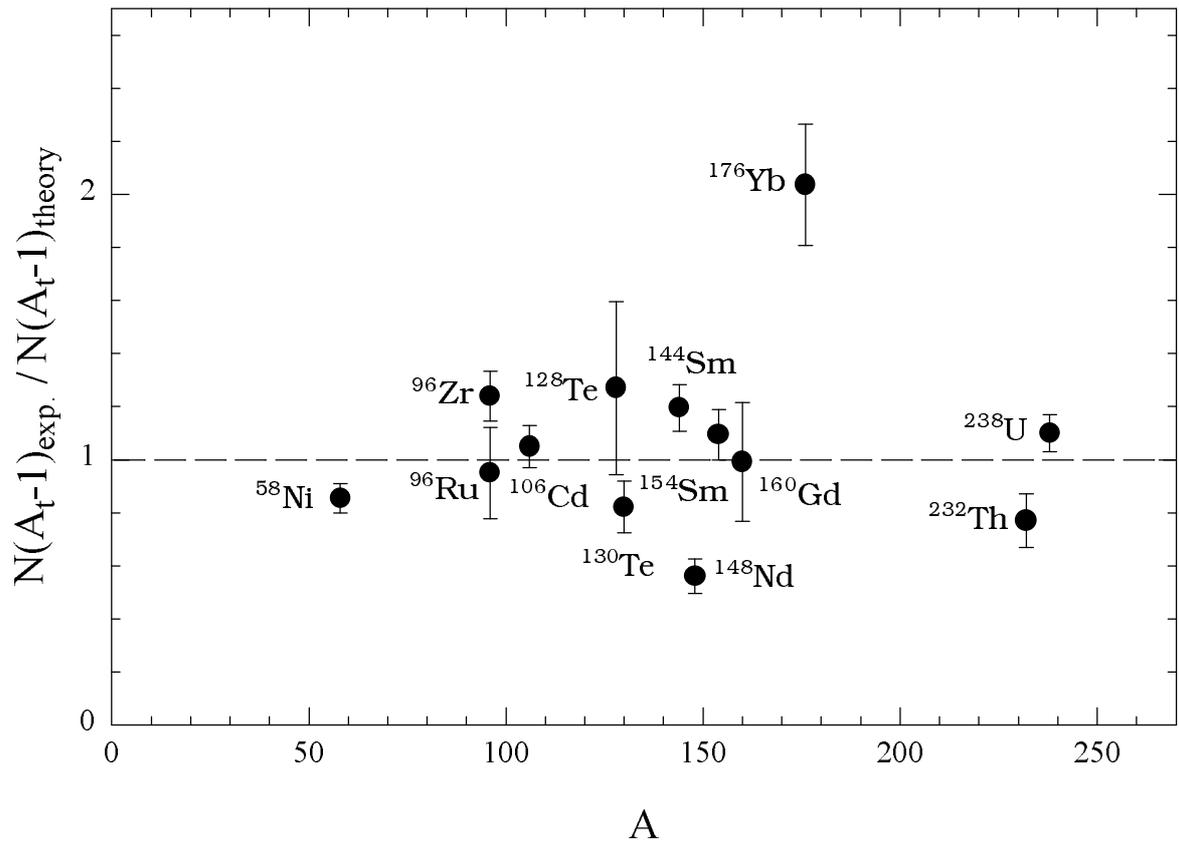
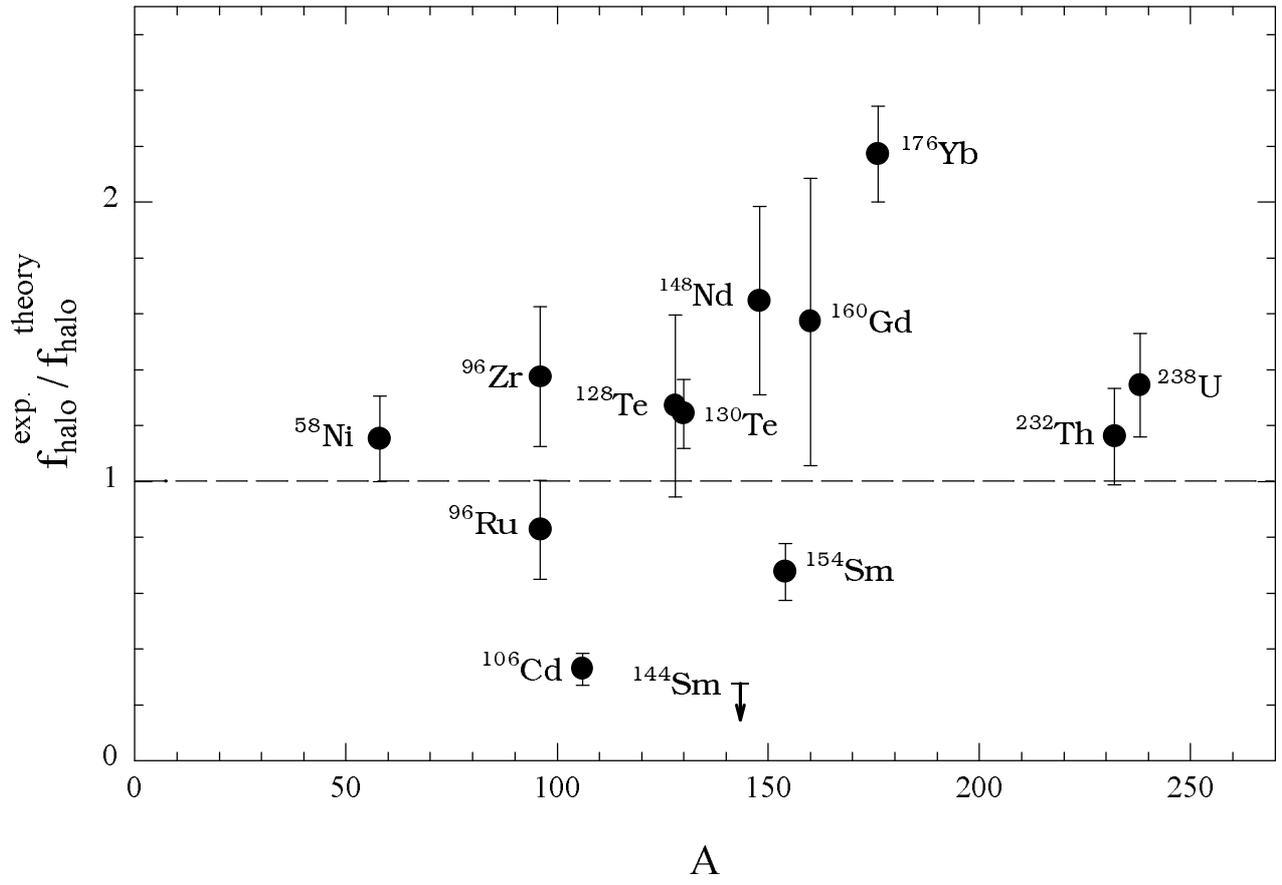
\end{document}